\documentclass[prb,twocolumn,amssymb]{revtex4}

\usepackage{graphicx}
\usepackage{dcolumn}
\usepackage{bm}

\begin{document}

\title{Resonancia ciclotr\'onica y dispersi\'on inel\'astica de luz\\ 
en puntos cu\'anticos semiconductores}
\author{Augusto Gonz\'alez}
\affiliation{Instituto de Cibern\'etica, Matem\'atica y F\'{\i}sica, Calle 
 E 309, Vedado, Ciudad Habana, Cuba}
\email{agonzale@icmf.inf.cu}
\author{Alain Delgado}
\affiliation{Centro de Aplicaciones Tecnol\'ogicas y
 Desarrollo Nuclear, Calle 30 No 502, Miramar, Ciudad Habana, Cuba}
\email{gran@ceaden.edu.cu}

\begin{abstract}
Resumimos un conjunto de resultados te\'oricos relacionados con la
absorci\'on en el infrarojo y la dispersi\'on inel\'astica (Raman) de luz
en puntos cu\'anticos semiconductores. Cuando es posible, se presenta la
comparaci\'on cualitativa con mediciones experimentales recientes.
\end{abstract}

\maketitle

\section{Introducci\'on}

El enorme inter\'es que se manifiesta recientemente hacia los dispositivos 
semiconductores de tama\~no nanom\'e\-tri\-co tiene dos aristas. Por un lado,
al reducir las dimensiones se puede aumentar el nivel de integraci\'on (es
decir, el n\'umero de elementos, transistores, etc por unidad de \'area). Y
por otro lado, al reducir las dimensiones dis\-mi\-nu\-yen los tiempos de 
respuesta, o sea que los dispositivos funcionan mas r\'apidamente. En ambos
casos, el resultado es un elemento electr\'onico u \'optico mas peque\~no y
potente.

La tendencia a reducir el tama\~no se encuentra, sin embargo, con un l\'{\i}mite
natural, el denominado l\'{\i}mite cu\'antico, que podr\'{\i}amos resumir en que 
los dispositivos no pueden ser mas peque\~nos que \'atomos, las cargas no mas 
peque\~nas que la del electr\'on, ni la intensidad de luz absorbida o emitida
menor que la de un fot\'on.

La nanotecnolog\'{\i}a y la nanociencia producen y estudian dispositivos
compuestos de varios \'atomos, o de varias mol\'eculas, o de varias capas 
at\'omicas en un s\'olido, o donde intervienen unos pocos electrones. Como
ejemplo citemos el trabajo [\onlinecite{Tarucha}] donde, a trav\'es de 
mediciones de conductancia de alta precisi\'on es posible establecer que
en un punto cu\'antico est\'an confinados 1, 2, 3, \dots, hasta 40 electrones.

En el presente art\'{\i}culo nos concentramos en procesos \'opticos que tienen
lugar en puntos cu\'anticos semiconductores, en particular la absorci\'on de 
luz en el infrarrojo y la dispersi\'on inel\'astica de luz (efecto Raman).
Aunque no se pretende una comparaci\'on directa con el expe\-ri\-mento, nuestros
resultados est\'an motivados por datos experimentales recientes \cite{Nickel,
Lockwood}, con los cuales se hace una comparaci\'on cualitativa. Los 
resultados son expuestos de forma resumida en las pr\'oximas secci\'ones. Los
detalles pueden hallarse en las versiones publicadas \cite{Ricardo,Alain1,
Alain2, Alain3}.

\section{Resonancia de ciclotr\'on en puntos cu\'anticos}

Consideremos un sistema de electrones cuyo movimiento est\'a confinado a un
plano y sobre los cuales act\'ua un campo magn\'etico perpendicular al plano
de movimiento. Tambi\'en a lo largo de la normal, hacemos incidir luz 
polarizada circularmente. La absorci\'on de luz en el infrarrojo mostrar\'a un
\'unico pico localizado en la denominada energ\'{\i}a de ciclotr\'on, $\hbar\omega_{ce}
=\hbar e B/m_e$. Este resultado se conoce como Teorema de Kohn \cite{Hawrylak}
y se debe a que la luz se acopla con el centro de masa del conjunto de 
electrones. Para un sistema de huecos, que se comportan como part\'{\i}culas con 
carga positiva, la posici\'on del pico de absorci\'on se halla por la misma
formula (basta sustituir $m_e$ por $m_h$), pero la po\-la\-ri\-zaci\'on de la luz
es opuesta. Los electrones absorben la polarizaci\'on $\sigma^+$ y los huecos
la $\sigma^-$.

Este razonamiento permite comprender a grandes rasgos el experimento 
reportado en [\onlinecite{Nickel}], donde en esencia se mide la absorci\'on de 
ciclotr\'on para electrones en un pozo cu\'antico como funci\'on de la concentraci\'on 
de electrones en el mismo. A diferencia del problema mencionado en el p\'arrafo
anterior, un segundo l\'aser con energ\'{\i}a superior a la brecha del semiconductor
crea pares electr\'on-hueco, con lo que el sistema no contiene s\'olo cargas
negativas, sino muchos electrones y un hueco. Por esa raz\'on, la posici\'on del
pico de absorci\'on deja de seguir la formula simple mencionada mas arriba y
comienza a depender de la concentraci\'on (densidad superficial) de electrones.
Los resultados experimentales son ``interpretados'' en t\'erminos de 
excitaci\'ones colectivas del tipo plasm\'on mas un hueco. Dichos resultados 
pueden resumirse en que el pico de absorci\'on se corre hacia el azul a medida 
que aumenta la concentraci\'on de electrones.

En el trabajo [\onlinecite{Ricardo}] se calcula la absorci\'on en el infrarrojo
de sistemas compuestos por unos pocos electrones (de 2 a 5) y un hueco. En el 
plano del movimiento se ha superpuesto, adem\'as, un potencial parab\'olico, con
el cual modelaremos un punto cu\'antico \cite{Hawrylak}. El trabajo extiende los
c\'alculos presentados previamente para el biexcit\'on (2 electrones mas 2 huecos)
en un punto \cite{biexciton}.

\begin{figure*}[t]
\begin{center}
\includegraphics[width=.6\linewidth,angle=-90]{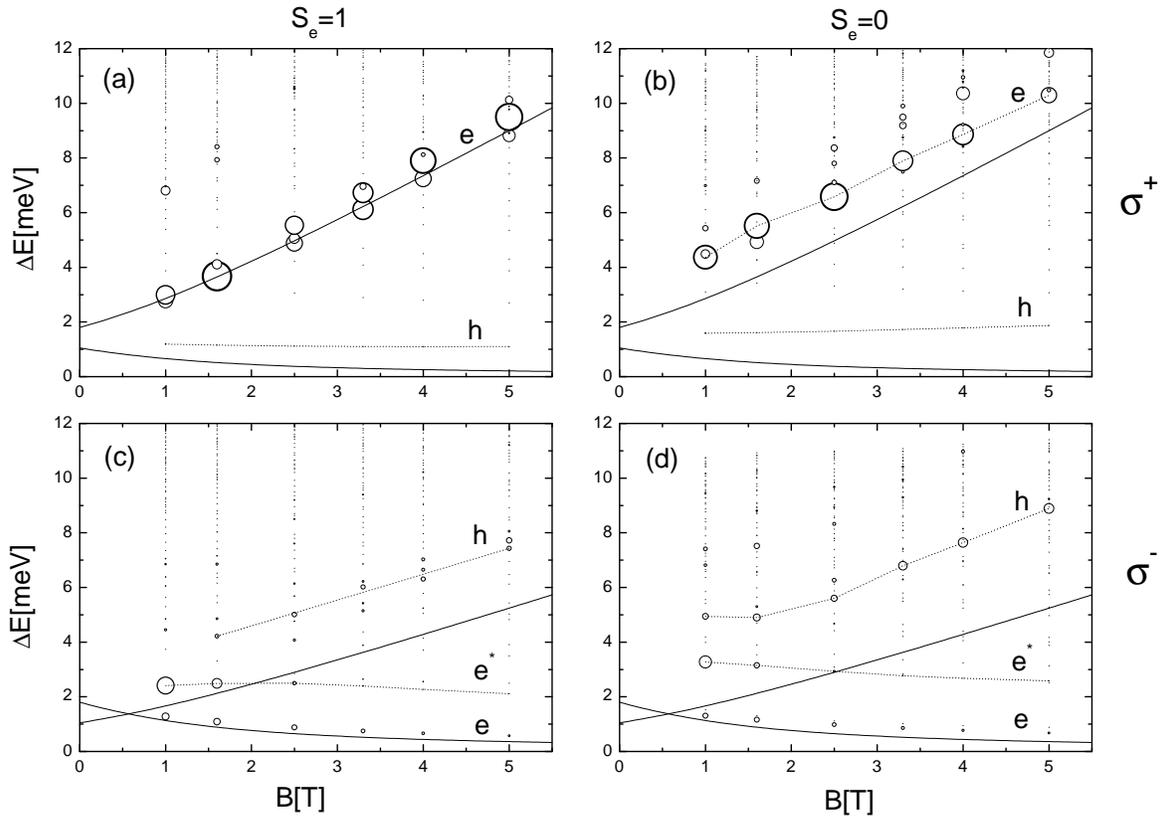}
\caption{\label{fig1} Posici\'on de los picos de absorci\'on y fortalezas de 
 oscilador para el tri\'on en estados tripletes ($S_e=1$) y singletes ($S_e=0$)
 de esp\'{\i}n y polarizaci\'on de la luz $\sigma^{\pm}$.} 
\end{center} 
\end{figure*}

La comparaci\'on directa de nuestros resultados con el experimento reportado en
[\onlinecite{Nickel}] desde luego que no es posible. El confinamiento lateral
puede inducir comportamientos que no est\'an presentes en el caso infinito. 
Sin embargo, uno espera que cualitativamente se reproduzca el corrimiento al
azul observado en el experimento. Por otro lado, no existen impedimentos para
realizar mediciones directamente en un punto cu\'antico o en un arreglo de puntos. 
N\'otese que en [\onlinecite{Nickel}] la absorci\'on en el infrarrojo se mide 
indirectamente a trav\'es de los cambios inducidos en la luminiscencia cuando el
haz infrarrojo es conectado. Y la luminiscencia es una t\'ecnica extremadamente
sensible y de facil implementaci\'on.

Para los c\'alculos utilizamos un modelo simple de dos bandas con par\'ametros
apropiados para el GaAs. La matr\'{\i}z hamiltoniana se diagonaliza exactamente en
una base de funciones de part\'{\i}culas libres en un campo magn\'etico. Las 
dimensiones de esta matr\'{\i}z est\'an entre 40,000 para los sistemas mas peque\~nos 
y 350,000 para los mas grandes. Matrices de estas dimensiones son 
diagona\-li\-zadas 
con ayuda del algoritmo de Lanczos \cite{RCF}, el cual nos permite obtener 
un conjunto de los autovalores mas bajos de energ\'{\i}a. El error estimado para
las energ\'{\i}as de excitaci\'on es de 0.02 meV y para las fortalezas de oscilador
(normalizadas a la unidad) de 0.02.

Un ejemplo de los resultados se muestra en la Fig. \ref{fig1}. En este caso, el 
sistema tratado es el tri\'on o $X^-$ ( es decir 2 electrones y un hueco, lo que 
dar\'{\i}a una carga neta igual a -1. En general, $X^{n-}$ designar\'a a un sistema 
compuesto por $n+1$ electrones y un hueco). Los gr\'aficos est\'an separados de 
acuerdo a la proyecci\'on sobre el campo magn\'etico del esp\'{\i}n total de los 
electrones, $S_e$ y a la polarizaci\'on de la luz absorbida. Los puntos en los
gr\'aficos dan las posici\'ones de los picos principales de absorci\'on. Como
informaci\'on adicional damos tambien la fortaleza de oscilador en forma de un
c\'{\i}rculo centrado en la posici\'on del pico de absorci\'on. De manera que un c\'{\i}rculo
grande simboliza un pico fuerte de absorci\'on.

En los gr\'aficos se ven tambi\'en l\'{\i}neas gruesas que simbolizan la posici\'on de los 
m\'aximos de Kohn para electrones o huecos en un punto cu\'antico parab\'olico
\cite{Hawrylak}:

\begin{equation}
\Delta E_{\pm}^{(e)}=\hbar\Omega_e\pm\frac{\hbar\omega_{ce}}{2},
\label{eq1}
\end{equation}

\begin{equation}
\Delta E_{\pm}^{(h)}=\hbar\Omega_h\mp\frac{\hbar\omega_{ch}}{2},
\label{eq2}
\end{equation}

\noindent
donde $\Omega_e=\sqrt{\omega_{0e}^2+\omega_{ce}^2/4}$, $\omega_{0e}$ es la
frecuencia corres\-pon\-diente al confinamiento lateral de los electrones, etc.
El sub\'{\i}ndice $\pm$ en la energ\'{\i}a se refiere a la absorci\'on de un fot\'on con
polarizaci\'on $\sigma^{\pm}$.

El resultado principal de nuestros c\'alculos se puede resumir en que la interacci\'on
entre los electrones y el hueco provoca un corrimiento de los picos de absorci\'on
con res\-pec\-to a la posici\'on predicha por (\ref{eq1}, \ref{eq2}). La dependencia
de este corrimiento respecto a la carga neta, $n$, del punto cu\'antico es la
siguiente: cuando $n$ va de 1 a 3 el corrimiento disminuye (es decir, los picos
se acercan a las posici\'ones dadas por (\ref{eq1}, \ref{eq2})), mientras que 
para $n=4$ crece nuevamente, manifestando la tendencia observada en el experimento 
de que cuando la concentraci\'on de electrones aumenta los picos de absorci\'on se
corren hacia el azul.

\section{dispersi\'on Raman en puntos cu\'anticos}

Consideremos, otra vez, el modelo simplificado de punto cu\'antico parab\'olico 
descrito en la secci\'on anterior, pero incluyamos en el decenas de electrones. La
intenci\'on es acercarnos a las condiciones de los experimentos 
[\onlinecite{Lockwood}], en los cuales se miden las intensidades Raman en puntos con
decenas o cientos de electrones. La dispersi\'on de la luz en este modelo se puede 
calcular a partir de la teor\'{\i}a de perturbaciones de segundo orden\cite{Loudon}.
Es decir, el tr\'ansito desde un estado inicial electr\'onico, $|i\rangle$, hasta un
estado final, $|f\rangle$, con la consiguiente absorci\'on de un fot\'on de frecuencia
$\nu_i$ y emisi\'on de otro de frecuencia $\nu_f$, se realiza a trav\'es del paso 
(virtual) por estados intermedios, $|int\rangle$. En las condiciones de los 
experimentos, la energ\'{\i}a del fot\'on incidente es $h\nu_i\gtrsim E_{gap}$, donde
$E_{gap}$ es la brecha del semiconductor, por lo que los estados intermedios 
contienen, adem\'as de los electrones iniciales, un par adicional electr\'on - hueco. 

La amplitud del proceso se calcula a partir de la expresi\'on:

\begin{widetext}
\begin{equation}
A_{fi}\sim \sum_{int} \frac{\langle f,N_i-1,1_f|H^+_{e-r}|int,N_i-1 \rangle
\langle int,N_i-1|H^-_{e-r}|i,N_i \rangle}{h\nu_i-(E_{int}-E_i)+i\Gamma_{int}},
\label{eq3}
\end{equation}
\end{widetext}

\noindent
donde $H_{e-r}$ es el hamiltoniano de interacci\'on de los electrones con la
radiaci\'on y $\Gamma_{int}$ es el ancho energ\'etico (fenomenol\'ogico) de los
niveles intermedios. $N_i$ es el n\'umero de fotones en el haz incidente.

El c\'alculo de $A_{fi}$ requiere de : a) Las energ\'{\i}as y funciones
de onda uniparticulares de electrones y huecos, que son obtenidas en la 
aproximaci\'on de Hartree-Fock. Las mismas son utilizadas como punto de partida 
en los pasos posteriores. b) Los estados finales de $N_e$ electrones, 
$|f\rangle$, hallados por medio de la aproximaci\'on de fase aleatoria (RPA 
por sus siglas en ingl\'es)\cite{nuclear}. c) Los estados intermedios de $N_e+1$ 
electrones y un hueco, $|int\rangle$, que se obtienen a partir del denominado 
formalismo pp-RPA. Y, finalmente, d) calcular los elementos de matr\'{\i}z de 
$H_{e-r}$ con estas funciones y realizar la suma (\ref{eq3}).

Con las amplitudes $A_{fi}$ uno calcula la secci\'on eficaz de dispersi\'on:

\begin{equation}
\frac{{\rm d}\sigma}{{\rm d}\nu_f}\sim \sum_f
|A_{fi}|^2 \delta (E_i+h\nu_i-E_f-h\nu_f), 
\label{eq4}
\end{equation}

\noindent
que es la magnitud medida experimentalmente.

Como se mencion\'o antes, en el conjunto de experimentos discutidos en 
[\onlinecite{Lockwood}] se obtienen los espectros Raman en puntos con
decenas o cientos de electrones. Mediciones sin o en presencia de campos 
magn\'eticos, sin o teniendo en cuenta la polarizaci\'on de la luz incidente
y reflejada, bajo diferentes condiciones de resonancia, son reportadas
en estos experimentos. De forma simplificada, podemos resumir los
resultados as\'{\i}: bajo condiciones de resonancia extrema (es decir
cuando $h \nu_i$ pr\'acticamente coincide con $E_{gap}$) el espectro
Raman es dominado por excitaciones uniparticulares, mientras que a 40 meV
o mas por encima de $E_{gap}$ los picos Raman est\'an asociados a 
estados finales que representan excitaci\'ones colectivas (de esp\'{\i}n o de
carga). En los experimentos, $h\nu_i$ no va mucho mas all\'a de 40 meV
por encima de $E_{gap}$ con el objetivo de no inducir procesos con
fonones \'opticos. El efecto principal del campo magn\'etico es mezclar
las excitaci\'ones de esp\'{\i}n y de carga, asi como desplazar la posici\'on de
los picos Raman.

\begin{figure*}[ht]
\begin{center}
\includegraphics[width=.65\linewidth,angle=-90]{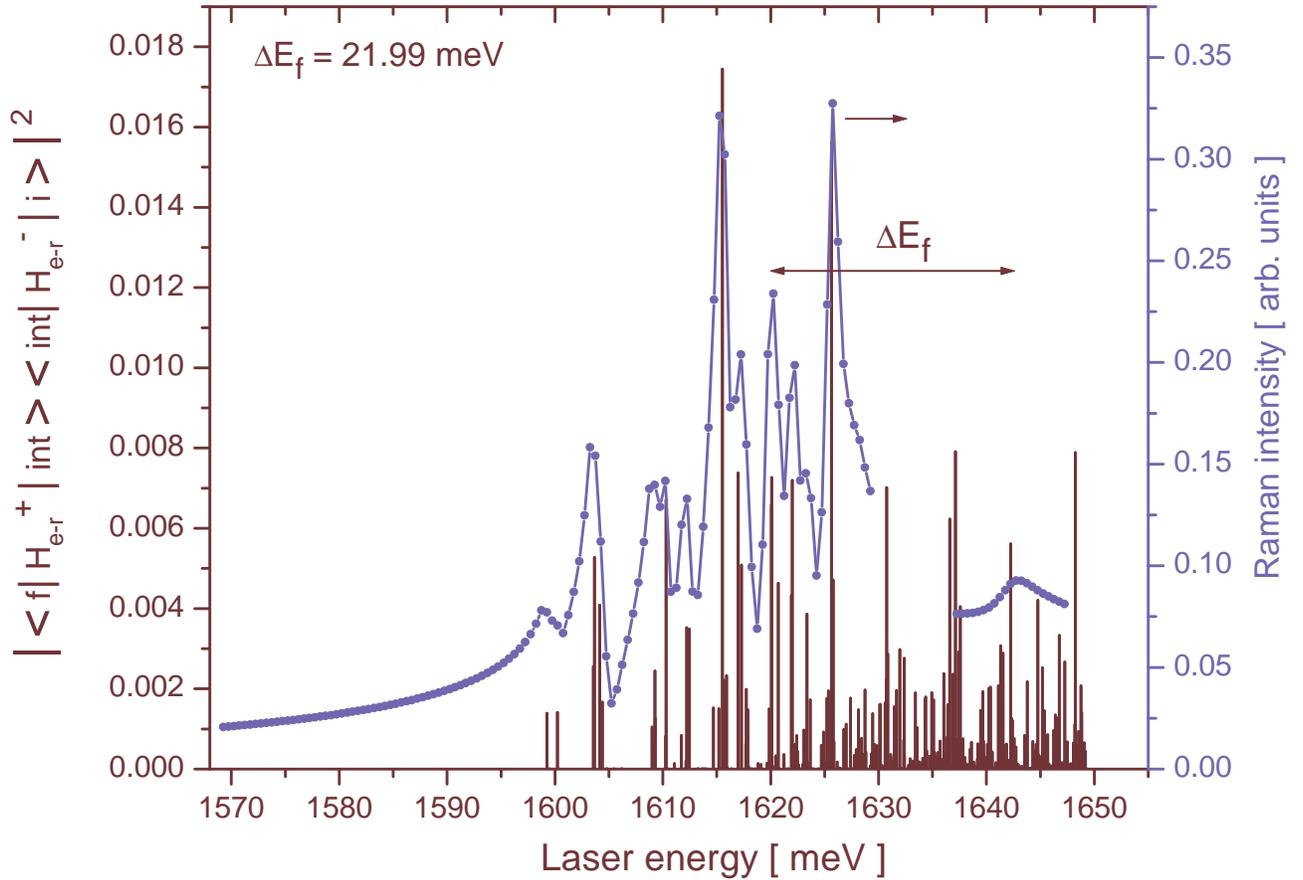}
\caption{Intensidad del pico Raman asociado a un estado final monopolar
 de carga como funci\'on de $h\nu_i$. Una resonancia en el canal de salida
 a 1642 meV es modelada. \label{fig2}}
\end{center}
\end{figure*}

En el art\'{\i}culo [\onlinecite{Alain3}] se presenta un an\'alisis detallado de la 
dispersi\'on Raman a cero campo magn\'etico. Se consi\-de\-ran tres condiciones de 
resonancia: a) $E_{gap}-30$ meV $< h\nu_i < E_{gap}$, b) $E_{gap} < 
h\nu_i < E_{gap}+30$ meV (resonancia extrema) y c) $E_{gap}+30$ meV $< 
h\nu_i$. El caso a) no ha sido abordado experimentalmente pero en el 
art\'{\i}culo se muestra su utilidad para identificar los diferentes picos en el
espectro. La teor\'{\i}a presentada en [\onlinecite{Alain3}] reproduce de forma
cualitativa los aspectos mas relevantes observados en el experimento y se\~nala
otros, como por ejemplo el papel de $\Gamma_{int}$ y su dependencia con la
energ\'{\i}a de excitaci\'on, que no hab\'{\i}an sido notados previamente.

Como ejemplo y resumen de los resultados de [\onlinecite{Alain3}], en la 
Fig. \ref{fig2} mostramos la intensidad del pico mas importante del espectro
(excitaci\'on monopolar colectiva de carga) como funci\'on de la energ\'{\i}a 
de excitaci\'on,  
$h\nu_i$. La brecha efectiva en este caso es aproximadamente 1600 meV.
En el caso $h\nu_i < E_{gap}$ observamos una dependencia mon\'otona de la 
intensidad del pico Raman. En resonancia extrema, por el contrario, las
oscilaciones abruptas de la intensidad est\'an asociadas a estados intermedios 
resonantes con $h\nu_i$. Cuando $h\nu_i > E_{gap}+30$ meV, el aumento 
de $\Gamma_{int}$ hace que se pierda la dependencia oscilante con 
$\Gamma_{int}$ observada antes y, en general, provoca una disminuci\'on de 
la intensidad. S\'olo para determinados estados intermedios del tipo 
``excit\'on + excitaci\'on colectiva'', $\Gamma_{int}$ conserva va\-lo\-res
relativamente peque\~nos. En la figura, uno de estos estados es el
responsable del aumento de intensidad en $h\nu_i\approx 1642$ meV.
En la figura se han incluido, adem\'as, en forma de l\'{\i}neas verticales los
numeradores que entran en la suma (\ref{eq3}). La comparaci\'on con las 
intensidades Raman muestra que los efectos de interferencia son poco
importantes en estos procesos.

Otros resultados con y sin campo magn\'etico externo inclu\'{\i}do son 
presentados en [\onlinecite{Alain2}]. El caso de puntos cu\'anticos neutros
o no dopados, donde una poblaci\'on de electrones y huecos con $N_e=N_h$
es inducida por un segundo l\'aser, es tratado en [\onlinecite{Alain1}].
Este \'ultimo caso hasta el momento no cuenta con una comprobaci\'on
experimental.

\section{Conclusiones y perspectivas}

Hemos presentado a grandes rasgos los resultados de los art\'{\i}culos
[\onlinecite{Ricardo,Alain1,Alain2,Alain3}] sobre absorci\'on en el
infrarrojo y dispersi\'on Raman en puntos cu\'anticos, as\'{\i} como los
ex\-pe\-ri\-mentos [\onlinecite{Nickel,Lockwood}] que motivaron estos
c\'alculos.

La perspectiva mas interesante que se abre ante nues\-tros ojos se
relaciona con el efecto Raman. C\'alculos deta\-lla\-dos en presencia de 
campo magn\'etico son interesantes de por si y necesarios en caso de
una comparaci\'on con resultados experimentales. Por otro lado, el
ancho de los niveles intermedios $\Gamma_{int}$ (o por lo menos la
contribuci\'on fon\'onica al mismo) puede ser calculado e introducido en
la expresi\'on (\ref{eq3}). Tal y como se mostr\'o en [\onlinecite{Alain3}]
con un ansatz fenomenol\'ogico para $\Gamma_{int}$, incluir la 
dependencia de $\Gamma_{int}$ con la energ\'{\i}a de excitaci\'on es 
imprescindible para una descripci\'on correcta de la dispersi\'on Raman
cuando $h\nu_i > E_{gap}+30$ meV. Otra posibilidad interesante es la
descripci\'on de las resonancias en los canales de salida, que en
[\onlinecite{Danan}] se interpretaron como ``excit\'on + excitaci\'on
colectiva''. La modificaci\'on requerida en nuestro forma\-lis\-mo 
consiste en
a\~nadir a los estados intermedios (que en esencia se construyen como un
par electr\'on - hueco por encima del estado de Hartree - Fock de $N_e$
electrones) otras componentes que correspondan a un par electr\'on - hueco 
mas una excitaci\'on del estado de Hartree - Fock. Estas pers\-pec\-tivas 
te\'oricas y el inter\'es mostrado por experimentalistas en retomar las 
mediciones Raman sobre bases superiores garantizan que el efecto Raman
con\-ti\-nua\-r\'a siendo un tema de trabajo excitante en los pr\'oximos a\~nos.

\vspace{.5cm}
\begin{center}
{\bf Agradecimientos}
\end{center}

Este art\'{\i}culo de revisi\'on est\'a basado en trabajos conjuntos con R. P\'erez, 
J. Mahecha, E. Men\'endez-Proupin y D.J. Lockwood.


\begin{thebibliography}{99}
\bibitem{Tarucha} S. Tarucha, D.G. Austing, T. Honda, Phys. Rev. Lett.
 {\bf 77}, 3613 (1996).
\bibitem{Nickel} H.A. Nickel, T.M. Yeo, A.B. Dzyubenko, B.D. McCombe, A. 
 Petrou, A.Yu. Sivachenko, W. Schaff, and V. Umansky, Phys. Rev. Lett. 
 {\bf 88}, 056801 (2002).
\bibitem{Lockwood} C.M. Sotomayor-Torres, D.J. Lockwood, and P.D.
 Wang, J. Electr. Materials {\bf 29}, 576 (2000).
\bibitem{Ricardo} R. P\'erez, A. Gonz\'alez and J. Mahecha, J. Phys.:
 Condens. Matter {\bf 15}, 7681 (2003).
\bibitem{Alain1} A. Delgado, A. Gonz\'alez and E. Men\'endez-Proupin, Phys.
 Rev. {\bf B 65}, 155306 (2002).
\bibitem{Alain2} A. Delgado and A. Gonz\'alez, J. Phys.: Condens. Matter
 {\bf 15}, 4259 (2003).
\bibitem{Alain3} A. Delgado, A. Gonz\'alez and D.J. Lockwood, 
 Phys. Rev. {\bf B 69}, 155314 (2004).
\bibitem{Hawrylak} L. Jacak, P. Hawrylak, and A. Wojs, {\it Quantum dots} 
 (Springer-Verlag, Berlin, 1998).
\bibitem{biexciton} R. P\'erez and A. Gonz\'alez, J. Phys.: Condens. Matter,
 {\bf 13}, L539 (2001).
\bibitem{RCF} Una descripci\'on breve del algoritmo de Lanczos puede 
 ha\-llar\-se en A. Gonz\'alez, Revista Cubana de F\'{\i}sica {\bf 19}, 5 (2002).
\bibitem{Loudon} R. Loudon, Adv. Phys. {\bf 13}, 423 (1964).
\bibitem{nuclear} P. Ring and P. Schuck, {\it The Nuclear Many-Body
 Problem} (Springer-Verlag, New-York, 1980).
\bibitem{Danan} G. Danan, A. Pinczuk, J.P. Valladares, L.N. Pfeiffer, K.W.
 West, and C.W. Tu, Phys. Rev. {\bf B 39}, 5512 (1989). 
\end{thebibliography}
\end{document}